\documentclass[aps,prc,twocolumn,showpacs]{revtex4}
\usepackage{amssymb}

\usepackage{amsmath,amssymb,mathrsfs}
\usepackage{graphicx}
\usepackage{amsthm}
\usepackage{amscd}
\usepackage{bm}
\usepackage{dsfont}

\newcommand\T{\rule{0pt}{2.8ex}}
\newcommand\B{\rule[-1.6ex]{0pt}{0pt}}

\usepackage{lscape}

\begin{document}

\title{Possibility of narrow resonances in nucleon-nucleon channels}
\author{M. I. Krivoruchenko}
\affiliation{Institute for Theoretical and Experimental Physics$\mathrm{,}$ B. Cheremushkinskaya 25 \\
117218 Moscow$\mathrm{,}$ Russia \\
Department of Nano-$\mathrm{,}$ Bio-$\mathrm{,}$ Information and Cognitive Technologies\\
Moscow Institute of Physics and Technology$\mathrm{,}$ 9 Institutskii per. \\
141700 Dolgoprudny$\mathrm{,}$ Moscow Region$\mathrm{,}$ Russia
}

\begin{abstract}

Compound states manifest themselves as bound states, resonances, or
primitives, and their character is determined by their interaction with the
continuum. If the interaction experiences a perturbation, a compound state
can change its manifestation. Phase analysis of nucleon-nucleon
scattering indicates the existence of primitives in the $^{3}S_{1}$, $%
^{1}S_{0}$, and $^{3}P_{0}$ channels. Electromagnetic interaction can shift
primitives from the unitary cut, turning them into narrow resonances. 
We evaluate this effect on the $^{1}S_{0}$ proton-proton scattering channel
in the framework of the Simonov-Dyson model. We show that electromagnetic interaction turns a
primitve with a mass of $ 2000$ MeV into a dibaryon resonance of approximately the same mass 
and a width of 260 keV. Narrow resonances of a similar nature may occur in other 
nucleon-nucleon channels. 
Experimental confirmation 
of the existence of narrow resonances 
would have important implications for the theory of nucleon-nucleon interaction.
\end{abstract}


\maketitle


Analytical properties of scattering amplitudes were extensively studied in
the 1950s, and for some time, the conditions of analyticity and unitarity were
considered likely to be sufficient for the full recovery of scattering
amplitudes. Low \cite{Low55} found the dispersion relation for scattering amplitudes, 
which takes into account analyticity and unitarity.
Solving this equation, Castillejo, Dalitz and Dyson \cite{CDD56} observed
an ambiguity, which is now called the CDD poles. The physical meaning of these
poles was clarified by Dyson \cite{Dys57}. Using a version of the Lee
model \cite{TDL54}, he showed that the CDD poles correspond to bound
states and resonances (for a review, see \cite{Shir69}).

$S$-matrices with a finite number of poles, which are mainly used in the
phenomenological parameterizations of scattering amplitudes, correspond to
zero-range singular potentials. The character of the interaction in such systems
is determined by analyzing the behavior of the scattering phase rather than
the potential: if the phase increases with energy, the interaction is an attraction;
if it decreases, the effect of interaction is a bound state or repulsion. According to the
Breit-Wigner formula, isolated resonances drive the phase shift up. In the
absence of bound states, 
the systems discussed in Refs. \cite{Low55,CDD56,Dys57} correspond to phases that increase with energy. 

In nucleon-nucleon interaction, the only bound state, the deuteron, arises in
the $^{3} S_{1}$ channel, whereas all other two-nucleon channels are free of
bound states.

Conversely, the nucleon-nucleon phase shifts decrease with increasing
energy and provide evidence for repulsion. The Dyson model \cite{Dys57}
was extended by Simonov \cite{Sim81} for the description of internucleon forces.
In the generalized model, both attraction and repulsion may dominate, and the
phase shifts behave bidirectionally. The repulsion has been included by
weakening the condition of strict positivity of the imaginary part of the $D$
function, thereby allowing zeros of the $D$ function to appear on the unitary
cut \cite{DFUN}.

Numerous experimental searches for exotic multiquark states did not give
decisive results. In the late '70s Jaffe and Low \cite{JLo79}
proposed an experimental method to identify exotic hadrons using a special formalism, called the $P$-matrix. 
The application of the $P$-matrix formalism to meson-meson scattering revealed $P$-matrix poles that
roughly correspond to the four-quark states predicted earlier by quark
models \cite{JLo79}. States that show up as poles of the $P$ matrix, rather
than the $S$ matrix, are called "primitives". They correspond to zeros of the $D$
function on the unitary cut, modify the Low
scattering equation, and generate the CDD poles \cite{MIK10}. 

The Simonov-Dyson model \cite{Sim81} provides a dynamical framework of the $P$-matrix formalism. 
It has been successfully applied to describe the nucleon-nucleon interaction
dominated by repulsion at short distances \cite{MIK10,KNSV85,BHGU85,FALE87,KNT11}. 
In the Simonov-Dyson model, primitives are the objects that produce repulsion.

The state of the art that prevailed in the early '80s can be characterized as
follows: The Dyson model describes systems dominated by attraction where bound
states and resonances may exist. The Simonov-Dyson model describes systems with
both attraction and repulsion where, in addition to bound states and resonances,
primitives come into play.

The conventional approach to the description of nucleon-nucleon interaction
is based on the Yukawa meson-exchange mechanism. As a result of the developments 
of the '70s and early '80s, a second approach has emerged, 
which ascribes nucleon-nucleon interaction to the $s$-channel exchange 
of $6q$-primitives (for a review, see \cite{NARO94}). 

The mechanism of $t$-channel meson exchange is studied in detail and is widely used to 
analyze particular effects. 
The $s$-channel exchange mechanism \cite{SCHA} has solid theoretical background and is also largely 
self-sufficient in modeling nucleon-nucleon interaction. 

The above two mechanisms may appear mutually exclusive. To decide which
of them is more appropriate, it is necessary to find a prediction that
could draw a clear distinction between the alternatives. The task is not
simple, because they both successfully describe a wide range of 
experimental data.

However, these mechanisms can also be dual to each other in the sense in which the $s$-channel resonance 
exchange is dual to the $t$-channel resonance exchange in the Veneziano model. In our case, this would mean that the $t$-channel 
meson exchange parameterizes the $s$-channel exchange of $6q$-primitives, and \textit{vice versa}. Simultaneous treatment 
of the $t$- and $s$-exchanges would represent in this case double counting.

There is, moreover, a third possibility. The hybrid Lee model can be reduced
to the Simonov-Dyson model \cite{MIK10}, so the latter can be regarded as a
phenomenological realization of the former. In the hybrid Lee model, 
the internucleon forces are generated by the $s$-channel exchange of $6q$-compound states.

We wish to clarify which of the aforementioned mechanisms is realized in nature, 
and whether there is a duality between 
the $t$-channel exchange of mesons and
the $s$-channel exchange of primitives ($6q$-compound states).

The existence of primitives imposes constraints on the interaction parameters. 
A characteristic feature of the Simonov-Dyson model is that small perturbations
of the interaction with the continuum shift primitives from the unitary cut
and, in the general case, turn them into resonances. 

Although primitives do not exist in the OBE models as elementary objects, the behavior of the phase 
shift associated with the repulsion permits a natural interpretation in terms of primitives. 
Any intersection of the phase shift of level $\delta (s)=0$ $\mod(\pi )$ with a negative slope can be
interpreted as a zero of the $D$ function and in turn as a primitive \cite{MIK10}. 
The status of primitives in the OBE models is roughly the same as the status of resonances and bound states.
Small perturbations of the $t$-channel exchange parameters move
primitives along the unitary cut. If there is a duality, the zeros of the $D$
function do not leave the unitary cut, so primitives remain primitives.
The duality condition thereby imposes constraints on the permissible perturbations of
the parameters of the Simonov-Dyson model, which determine, in particular, compound state masses and 
their coupling with the continuum.

A similar situation occurs in the hybrid Lee model. The Simonov-Dyson
model ensures the existence of stable primitives, if it arose as a reduction
of the hybrid Lee model. 
In this case, arbitrary perturbations of the parameters of the hybrid Lee model are allowed, 
whereas the corresponding variations of parameters of the Simonov-Dyson model are those 
that keep primitives on the unitary cut.

Some discrimination between the aforementioned alternatives can be made, if
one considers scattering in an external field. In the OBE
models and in the case of duality, the scattering cross section will not be
qualitatively changed. Conversely, if there is no duality and the $s$-channel
exchange mechanism operates as assumed in the Simonov-Dyson model we can
expect the appearance of a resonance in the place where we earlier observed
a primitive. Because the mass of the resonance is approximately known 
(it corresponds to zero phase shift), the problem reduces to the estimation of the width of the resonance.
In the hybrid model of Lee, primitives are rigidly connected with the unitary cut,
so the resonance will not occur.

In fact, we can proceed without external fields. 
A meaningful example of the perturbation of strong-interaction parameters of the model
is given by the electromagnetic interaction.
In this case, the condition for the existence of primitives is violated as a result of electromagnetic 
mass shifts of the nucleons and the $6q$-compound states, as well as modification of the imaginary part 
of the $D$ function, which additionally receives the Gamow-Sommerfeld factor for the proton-proton channel. Inclusion
of the electromagnetic effects allows the determination of the width of the resonance, as
well as the (small) shift of the resonance mass relative to the primitive mass.

We therefore consider the possibility that the displacement of primitives is associated with
electromagnetic interaction. This takes place in all nucleon-nucleon channels 
because of the electromagnetic mass shifts and other electromagnetic corrections, 
including the initial- and final-state Coulomb
interaction. For numerical estimates, we use the Simonov-Dyson model \cite{comment2}.

The $D$ function can be written as
\begin{equation}  \label{GEQ110}
D(s)=\Lambda (s)-\Pi (s),
\end{equation}
where
\begin{eqnarray}
\Lambda ^{-1}(s) &=& \sum_{\beta }\frac{g_{\beta }^{2}}{s-M_{\beta }^{2}} + f, \label{LAMB} \\
\Pi (s)          &=& -\frac{1}{\pi }\int_{s_{0}}^{+\infty }\Phi
_{2}(s^{\prime })\frac{\mathcal{F}^{2}(s^{\prime })}{s^{\prime }-s}ds^{\prime }.
\label{SELF}
\end{eqnarray}
The indices $\beta $ label $6q$-compound states 
with masses 
$M_{\beta } $, the coupling constants $g_{\beta } $ parametrize the interaction
with the continuum, and $f$ describes the four-fermion contact interaction. 
$\Phi _{2} (s)=\pi k/\sqrt{s} $ is the phase space
volume of two nucleons, where $k$ is the nucleon momentum in the center-of-mass frame. 
$\mathcal{F}(s)$ is
the form factor arising from the vertices of compound states and nucleons 
and the four-fermion interaction vertices:
\begin{equation}
\mathcal{F}(s)=\left( \frac{s}{s_{0}}\right) ^{1/4}\frac{\sin (kb)}{kb}C_{0}(s).  
\label{GEQ113}
\end{equation}
Here, $s_{0}$ is threshold value of $s$ and $b$ is the
effective interaction radius. The Gamow-Sommerfeld factor \cite{Gam28,Som39}
\begin{equation}
C_{0}^{2}(k)=\frac{2\pi \eta }{\exp (2\pi \eta )-1},
\end{equation}
where $\eta =\alpha \mu /k$ and $\mu =m/2$ is the reduced proton mass,
accounts for the initial- and final-state Coulomb interaction of the $S$-wave protons. 
In the $pn$ and $nn$ channels, $C_{0}(k)=1$. 

In the proton-proton channel, the dispersion integral of Eq.~(\ref{SELF})  
can be evaluated to give
\begin{widetext}
\begin{eqnarray}
\kappa \Pi (s) &=& \frac{1}{2k^{2}b}\left( 2\mathrm{Si}(2kb) \sin (2kb)+(\mathrm{Ci}(2kb)+\mathrm{Ci}(-2kb))\cos (2kb)-2C-\ln (-4k^{2}b^{2})\right) 
\nonumber \\
                  &-& \frac{\sin (kb)}{kb}e^{ikb}\left( C_{0}^{2}(k)+\frac{\pi }{k}\right) 
                   + \sum_{n=1}^{\infty }\frac{\sinh (b/n)}{b/n}e^{-b/n}\frac{2}{1+k^{2}n^{2}},
\label{GEQ114}
\end{eqnarray}
\end{widetext}
where $\kappa =2mb/\pi$ and $C=0.5772\ldots $ is the Euler constant.
Here units are chosen in which $a_c \equiv 1/(\alpha \mu)  = 1$, later we will restore the dimension. 
In the expression (\ref{GEQ114}), 
a branch of the logarithm $\ln (1)=0$ enters; in the complex $k$-plane, the cut extends 
from $-\infty $ to $0$, so $\ln (-1\pm i0)=\pm i\pi $. In the physical upper half of 
the complex $k$-plane, the self-energy operator $\Pi (s)$ is an analytical function. 
On the real $s$-axis below $s_0$ or, equivalently, on the imaginary half-axis $\Re k=0$, $\Im k>0$, 
$\Pi (s)$ is a real function. 

$\Lambda ^{-1} (s)$ is assumed to have a single pole at $s=M^{2} $, corresponding
to a $6q$-compound state. The contribution to $\Lambda ^{-1} (s)$ of the contact four-fermion 
interaction is described by a constant term. As a result,
\begin{equation}
\kappa ^{-1}\Lambda ^{-1}(s)=c_{p}(\frac{r_{p}}{s-M^{2}}-\frac{r_{p}}{%
s_0-M^{2}})-\frac{1}{\gamma },  
\label{GEQ115}
\end{equation}
where $\kappa c_{p}r_{p}=g_{1}^{2}$ is the coupling constant entering Eq.~(\ref{LAMB}) 
and $r_{p}=8\pi ^{2}/b^{2}$ is the residue
of the free $P$ matrix. 
The value of $c_{p}$ measures the strength of interaction of the compound state 
with the continuum.
The residue of $\kappa ^{-1}\Lambda ^{-1} (s)$ 
is parameterized as follows
\[
c_{p}r_{p} = \xi\frac{M^{2} - s_0}{\gamma }.
\]
In the $pn$ and $nn$ channels, the $D$ function has the correct analytical properties,
provided $\xi \in (0,1)$. The phase
shifts are well reproduced for $\xi = 0.9$. 
Here we restricted ourselves to verification of the fact that, 
in the channel $pp$ at $\xi = 0.9$, the analytical
properties of the $D$ function are correct. 

The value of $\gamma $ is determined by the scattering length. 
In the $pp$ channel, we use the series expansion around $k=0$ 
for $\Re k > 0$
\cite{LALI}:
\begin{equation}
kC_{0}^{2}(s)\cot \delta (k)+2h(k)=\frac{1}{a}+\frac{r}{2}k^{2}+...
\end{equation}
This expansion allows one to draw a link between the low-energy parameters of the model and
the phenomenological constants that characterize the low-energy proton-proton scattering. 
The following equation serves as the definition of an analytical function $h (k)$:
\begin{equation}
2\psi (1+\frac{i}{k})+ik+\ln (-k^{2})=2h(k)+ikC_{0}^{2}(s),
\end{equation}
with $\psi(z)$ being the digamma function.

The strong-interaction phase shift can be found from the strong-interaction $S$ matrix 
modified by the initial- and final-state Coulomb interaction
\begin{equation}
S^{\prime}(k) \equiv e^{ 2i\delta (k)} =\frac{D(s-i0)}{D(s+i0)}.
\end{equation}

We use the empirical value of the scattering length $a$ to fix $\gamma $:
\begin{eqnarray}
\gamma  =  1&+&\frac{b}{a}+\left( \ln \left( 4b^{2}\right) +4C-3\right) b \nonumber \\
       &-&\sum _{m=2}^{\infty }(-1)^{m} \frac{2^{m+1} }{(m+1)!} b^{m} \zeta (m) ,
\end{eqnarray}
where $\zeta (m)$ is the Riemann zeta function. Once the model parameters are determined, 
we arrive at a prediction for the effective range
\begin{eqnarray}
r &=& \frac{2}{3} \left(1+\gamma \right)b-\frac{8\gamma \xi}{b\, \left(M^{2} -4 m^{2} \right)} 
-\frac{7}{9} b^{2} \\
 &+& \sum _{m=2}^{\infty }(-1)^{m} \frac{2^{m+2} (m-1)(m+6)}{3(m+3)!} b^{m + 1} \zeta (m). 
\nonumber 
\end{eqnarray}
In the $^1 S_0$ $pn$ channel with $a_c = \infty$, one has $r = 2.1$ fm \cite{MIK10}, 
which should be compared with $r = 2.8$ fm \cite{NAG79} in the OBE models. 

In bare strong interaction, the real and imaginary parts of the $D$ function simultaneously vanish 
at some point on the real $k$-axis. 
The zeros of $\Im D (s)$, on the other hand, are the zeros of $\mathcal{F}(s)$, 
the lowest-lying one is located at \cite{comment}
\begin{equation}
k^*_{0} = {\pi}/{b_0}.
\label{ft}
\end{equation}
In the parameterization (\ref{GEQ115}), $\Re D(s)$ also vanishes at this point, whereas 
the compound state mass is equal to the total energy of protons in the center-of-mass frame:
\begin{equation}
M_{0} = 2\sqrt{\pi ^{2} /b_{0} ^{2} + m_{0} ^{2} }.
\label{M0} 
\end{equation}

In strong plus electromagnetic interaction (strong interaction + QED), 
the zero of the $D$ function is shifted to the complex plane. 
The effect is related to the change of the model parameters 
$b_0 \to b = b_0 +\Delta b $, $m_0 \to m = m_0 +\Delta m $, etc., 
and to the Coulomb interaction of the protons. 
Let us discuss the change of parameters.

Electromagnetic mass splitting of hadrons is well studied. QED corrections to hadron masses
are usually attributed to the Coulomb interaction between the quarks and their spin-spin interaction:
\begin{equation}
\Delta \mathcal{M} = c < \sum_{i<j} e_{i}e_{j} >
                      + h < \sum_{i<j} e_{i}e_{j} \mbox{\boldmath{$\sigma$}}_{i} \mbox{\boldmath{$\sigma$}}_{j} >.
\label{pb}
\end{equation}
Here, it is assumed that all the quarks are in $S$-wave states. The averaging is made over the color-spin-isospin wave function
of the quarks. A reasonable description of the mass splitting in octet-baryon isomultiplets, accounting 
for the mass difference of nonstrange quarks, can be obtained for $c = 3.06$ MeV and $h = - 1.35$ MeV \cite{Ros98}. 

The average values of the spin-isospin operators and the corresponding
electromagnetic mass shifts of the proton, the neutron, and the $6q$-compound states
with quantum numbers of the $^1 S_0$ nucleon-nucleon channels are listed in Table \ref{tab}.
The calculation of average values of the operators was carried out 
using fully antisymmetrized color-spin-isospin quark wave functions. 
The resonance is denoted by $d^*$.

For the nonstrange quarks, the MIT bag model gives $b \sim R$ and $M \sim R^3$, 
where $R$ is the radius of the 6-quark bag, and thus the change of the parameter $b$ for the inclusion 
of electromagnetic interaction is simply given by
\begin{equation}
\frac{\Delta b }{b} =\frac{\Delta M }{3M}.
\label{dbob}
\end{equation}

The values of the $^1 S_0$ scattering lengths, derived from the experimental data, 
are as follows \cite{NAG79}: 
$ 7.83 \pm 0.01 $ fm, $ 23.75 \pm 0.01 $ fm, and $ 16.4 \pm  0.09$ fm in the $ I_3 = 1,0, -1 $  
channels respectively. 
In our approach, the scattering length or, equivalently, the value of $\gamma$ is adjustable 
parameter. We obtain $a = 6.5$ fm in the $pp$ channel by fitting a maximum of the phase 
at $ T \approx 10 $ MeV. 

To compare the physical phase with that of the bare strong interaction, we must have an estimate of $ a_0 $. 
We give some heuristic arguments in favor of the change in the scattering length 
being proportional to the electromagnetic mass shift of the $6q$-compound state 
with quantum numbers of the channel.


\begin{table} [!ht]
\centering
\renewcommand{\arraystretch}{1.1}
\caption{
Electromagnetic mass shifts of the proton, the neutron, and the $6q$-compound states $d^{*}$ 
with isospin projections $I_3 = \pm 1,0$. The third and fourth columns show average values of the spin-isospin operators. 
Electromagnetic mass shifts of the hadrons are shown in the last column (in MeV). 
}
\vspace{1mm}
\label{tab}
\begin{tabular}{|lrrrr|}
\hline
\hline
 Hadron  \T \B            &   $I_3$                & $< \underset{i<j} \sum e_{i}e_{j} >$ & $<\underset{i<j} \sum e_{i}e_{j} \mbox{\boldmath{$\sigma$}}_{i} \mbox{\boldmath{$\sigma$}}_{j} >$ & $\Delta \mathcal{M}$ \\ \hline
 $~~~~~p$                 &   $\frac{1}{2}$        & $0~~~~~~$                           & $\frac{4}{3}~~~~~~~~~~$                                                        & -1.8~~                     \\
 $~~~~~n$                 &   $-\frac{1}{2}$           & $-\frac{1}{3}~~~~~~$              & $1~~~~~~~~~~$                                                                    & -2.4~~                     \\
 $~~~~~d^{*++}$           &   $1$                  & $1~~~~~~$                           & $-\frac{7}{5}~~~~~~~~~~$                                                       &  5.0~~                     \\
 $~~~~~d^{*+}$            &   $0$                  & $-\frac{1}{3}~~~~~~$              & $\frac{19}{5}~~~~~~~~~~$                                                       & -6.2~~                    \\
 $~~~~~d^{*0} $  \B       &   $-1$                 & $\frac{2}{3}~~~~~~$              & $-\frac{2}{5}~~~~~~~~~~$                                                       & -1.5~~                     \\ 
\hline
\hline
\end{tabular}
\end{table}


In the Born approximation, the scattering length is proportional to the averaged interaction potential. 
A part of this potential associated with the electromagnetic interaction leads to a difference 
of scattering lengths in different isospin channels.
On the other hand, the electromagnetic mass splitting of hadrons is also determined by averaging 
the electromagnetic potential, although of quarks rather than nucleons. 
In any case, the equation 
$ a = a_{0} + C_{2} \Delta M$, with $ a_0 = 14.45 $ fm and $ C_2 = - 1.45 $ fm/MeV, gives 
scattering lengths $a = $ 7.3 fm, 23.4 fm, and 16.6 fm in the $ I_3 = 1,0, -1 $ channels, respectively, 
which agrees well with the empirical data. Thus, we accept 
\begin{equation}
\Delta a =C_{2} \Delta M . 
\end{equation}
A slightly better fit of the proton-proton phase shift is obtained for $a = 6.5$ fm. 
Because $a/a_c \sim 1/3$ (in dimensional units $a_c = 58$ fm), the electromagnetic corrections to low-energy
parameters of the model are expected to be only moderately small. In the scattering lengths, 
the electromagnetic corrections  $\sim 1$.

Parameter $\xi$ is also an adjustable parameter. We will accept $\Delta \xi = 0$. 

We thus have estimates of the proton mass $m_0$ in the bare strong interaction, 
as well as the electromagnetic mass corrections for 
nucleons and $d^{*}$. The relative value of the correction to $b$ is known also.

The proton kinetic energy $T_{\delta}= 244$ MeV in the laboratory frame corresponds to a vanishing phase shift. 
In the center-of-mass frame, one finds the momentum of the protons $k_{\delta} =\sqrt{mT_{\delta} /2} $ and
their total energy $M_{\delta} = \sqrt{2 mT_{\delta} + 4 m^{2}}  = 1995$ MeV.
The phase shift vanishes provided that the imaginary part of the $D$ function vanishes;
that is, it is necessary to have $k_{\delta} b = \pi $, so the interaction radius $b$ is fixed. 
The total mass of the compound state $M \approx M_{\delta} $ is known with an accuracy of $O( \Delta M)$, 
so we find from Eq.~(\ref{dbob}) the interaction radius $b_0$ with electromagnetic interaction switched off. 

The condition (\ref{ft}) holds for bare strong interaction. Using Eq.~(\ref{M0}) one may find a mass of the compound 
state with electromagnetic interaction switched off. Then, for $\alpha \neq 0$, 
the mass $M = M_{0} +\Delta M$.

\begin{center}
\begin{figure}[h] 
\includegraphics[angle = 0,width=0.450\textwidth]{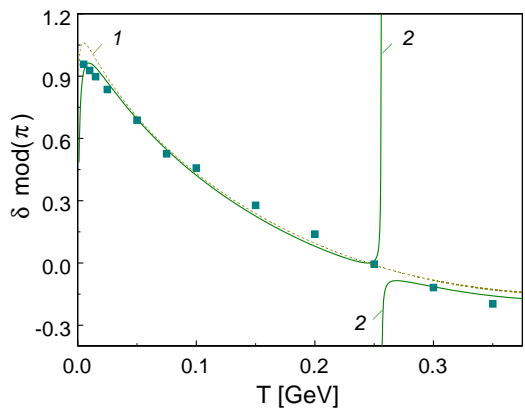}
\caption{(color online)
The $^1 S_0$ proton-proton scattering phase shift in radians modulo $\pi$ versus the proton kinetic energy 
in the laboratory frame. 
The smooth dashed curve \textit{1} is the model prediction with electromagnetic interaction switched off. 
The solid broken curve \textit{2} arises 
upon dressing the model parameters by electromagnetic interaction, i.e., in strong interaction + QED. 
The squares show the experimental data
\cite{PHAS}.
}
\label{fig1}
\end{figure}
\end{center}

\begin{center}
\begin{figure}[h] 
\includegraphics[angle = 0,width=0.450\textwidth]{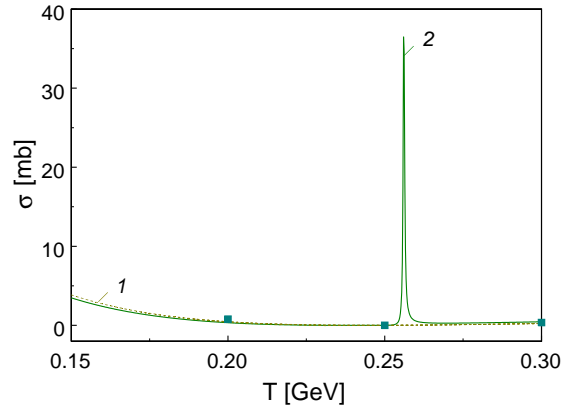}
\caption{(color online)
The $^1 S_0$ proton-proton partial-wave cross section versus the proton kinetic energy 
in the laboratory frame. 
Notations are the same as in Fig.~\ref{fig1}.
}
\label{fig2}
\end{figure}
\end{center}

In Fig.~\ref{fig1}, the proton-proton scattering phase shift is plotted in strong interaction + QED 
and in the bare strong interaction case. In Fig.~\ref{fig2}, the $S$-wave proton-proton cross section is plotted. 
The narrow resonance peak is associated with the complex root of the equation
\begin{equation}
D(M^2_{*} - iM_{*} \Gamma_{*}) = 0.
\label{droot}
\end{equation}
If the electromagnetic interaction is switched off, Eq.~(\ref{droot}) has a primitive root 
corresponding to $\Gamma_{*} = 0$.
When the electromagnetic interaction is switched on, a resonance of a width $\Gamma_{*} \neq 0$ appears
in the neighborhood of the primitive. 

The full set of the model parameters is shown in Table \ref{tab2}. 
In the first row, the model parameters to which we previously 
attributed the subscript $0$ are given.

The value of $M_{\delta}$ is defined to within a few MeV. If we take $M_{\delta} = 2000$ MeV, 
the resonance will have a mass of $M_{*} = 2005.5$ MeV and a width of $\Gamma_{*} = 250$ keV. 
Because the position of the peak is not very well defined, in an experimental search for the resonance, 
one needs to scan a region around the zero value of the scattering phase
with a resolution in $\sqrt{s}$ better than 100 keV. 

Phase analysis of nucleon-nucleon scattering did not reveal
primitives in the $^{3}P_{1}$ and $^{1}P_{1}$ channels. In the $^{3}P_{0}$ channel, 
the data indicate the existence of a primitive with a mass of about 1970
MeV \cite{KNT11}. Arguments similar to those used for the $S$-wave scattering
allow one to conclude that electromagnetic interaction in the $^{3}P_{0}$
proton-proton channel transforms the primitive to a resonance with a mass of about 1970 MeV
and a width of a few hundred keV.

\vspace{-6mm}
\vspace{-6mm}

In nuclear matter, nucleons are in an effective mean field. Here we should
expect significant renormalization of the parameters responsible for the
dynamics. In the Simonov-Dyson model, we also should expect the transformation of the
$6q$-primitives to the resonances, which significantly modifies
the properties of nuclear matter. Dibaryon resonances are bosons. In cold
nuclear matter and with increasing chemical potential of nucleons, a Bose
condensation will begin \cite{Bal84}. This modifies the equation of state (EoS)
of nuclear matter and modifies the properties of neutron 
stars under equilibrium \cite{MIK87}. Bose condensation of dibaryon resonances may
also be the trigger for a phase transition to quark matter. The above
scenario and its connection with other effects in nuclear physics and
astrophysics are discussed in Ref. \cite{KNT11}.

\vspace{-6mm}
\vspace{-6mm}


Astrophysical observations of neutron stars provide important constraints 
on the nuclear matter properties above saturation density and, indirectly,
on the in-medium modifications of hadrons. 
The problem of the possible conversion of primitives to resonances, however, 
can most reliably be settled in the laboratory.

In the phase behavior of the nucleon-nucleon scattering, there are clear features that
can be attributed to primitives. We therefore believe that in bare strong interaction
some compound states remain on the unitary cut and manifest 
themselves as primitives, rather than resonances or bound states. When external conditions, 
such as non-zero density of matter or finite temperature, become different or new interactions 
come into play, compound states can leave the unitary cut, turning into resonances.
These resonances can be observed in the laboratory as the usual Breit-Wigner peaks.

There is no conclusive evidence for the existence of exotic $6q$ resonances. However, in the phase analysis of $NN$ scattering $6q$ primitives are determined reliably. 
In the $P$-matrix formalism primitives are identified with exotic multiquark states. 
If primitives are shifted from the unitary cut under influence of perturbations, 
they can be observed as narrow resonances.

In the OBE models and the hybrid Lee model primitives remain on the unitary cut. 
In the Simonov-Dyson model, in general, the primitives, if exist, are shifted from the unitary cut. 
Some preferences for these models can be done after the search results of the narrow dibaryons will become available.

For an experimental search for narrow dibaryon resonances, one should have a
beam of protons with a kinetic energy of $T \sim 250$ MeV and an energy
spread below 100 keV. The energy resolution of the detector is not important, 
so one can use scintillation detectors. In the CELSIUS
accelerator at Uppsala, the beam momentum spread
was a few times $10^{-3}$ before electron cooling and
a few times $10^{-5} $ after electron cooling. Under such conditions, 
it is possible to measure $\sqrt{s}$ 
with an accuracy of $10$ keV or better.
The narrow width of $d^{*++}(2000)$ is not an obstacle to its experimental search.

Resonances of the same nature should exist in the $pn$ and $nn$ channels. 
Experiments with neutron beams are, however, more complicated. Also, e.g., deuterium cannot be used as a fixed target, 
because the narrow resonances are smeared out by the Fermi motion.
We thus estimated the effect in the proton-proton channel, which is of interest for experiments 
with protons bombarding a hydrogen target.


\begin{table} [t]
\addtolength{\tabcolsep}{2pt}
\centering
\renewcommand{\arraystretch}{1.1}
\caption{
Parameters of the Simonov-Dyson model for $M_{\delta} = 1995$ MeV 
with electromagnetic interaction switched off and on. $a$ is the scattering length, $r$ is the effective range, 
and $b$ is the interaction radius (in fm). $m$ is the proton mass, 
$M$ is the compound state mass, and
$M_{*}$ and $\Gamma_{*}$ are the primitive 
and resonance masses and widths (in MeV).
}
\vspace{1mm}
\label{tab2}
\begin{tabular}{|cccccccc|}
\hline
\hline
$\alpha$   \B     & $a$ & $r$ & $b$ & $m$   & $M$    & $M_{*}$ & $\Gamma _{*}$  \\ 
\hline
   0            & 14.45                  & 2.22                   & 1.829                  & 940.1 & 1998.6 & 1998.6  & 0              \\ 
${1}/{137}$     & 6.50                   & 0.84                   & 1.831                  & 938.3 & 2003.5 & 2000.5  & 0.260          \\ 
\hline
\hline
\end{tabular}
\end{table}


Suppose that, in proton-proton scattering, experimentalists scanned the energy region 
up to a few MeV around the value where the phase shift is zero and the resonance was not found.
In this case, we would have to conclude that the primitives are stable under perturbations 
and are strictly adhered to the unitary cut.

Such stability is inherent in the OBE models and hybrid Lee model. 
Consequently, the Simonov-Dyson model would either be dual to the OBE models or 
be a phenomenological realization of the hybrid Lee model.

Assume instead that, in place of the primitive, experimentalists found the $d^{*++}(2000)$ resonance with a width of about 260 keV.
This would mean that the primitives are not strictly adhered 
to the unitary cut. This property holds only in the Simonov-Dyson model.

Observation of the resonance will have the following consequences: 
i) mutual transformation of primitives and resonances will be confirmed, 
ii) universality of the Yukawa meson-exchange mechanism will be questioned on the experimental basis, 
iii) the region of validity of the hybrid Lee model will be reduced, and
iv) the absence of duality between the $t$-channel meson exchange and the $s$-channel exchange of $6q$-primitives will be proved.

\vspace{5mm}

The author is grateful to M.~G.~Schepkin and Yu.~A.~Simonov for useful discussions. 
This work is supported in part by RFBR grant No. 09-02-91341 and grant of Scientific Schools of Russian Federation No.
4568.2008.2.



\begin{thebibliography}{99}
\bibitem{SCHA} In terms of crossing, all channels are equivalent, 
while the single-particle propagators that link blocks of strongly connected diagrams 
describe the "exchange". We thus find it possible to talk about the $s$-channel exchange of $6q$ primitives, 
since this stresses crossing symmetry of the amplitude.
\bibitem{comment2} This model is also known as the Quark Compound Bag model.
\bibitem{comment} Parameters of the model with electromagnetic interaction switched off, i.e., 
in bare strong interaction, are supplied with the subscript $0$.
\end{thebibliography}
\end{document}